\documentclass[aps,pre,showpacs,amssymb,superscriptaddress,nofootinbib,twocolumn]{revtex4-1}

\usepackage{amsmath,amsfonts,amsthm,amssymb}
\usepackage{braket}
\usepackage{setspace}
\usepackage{amsmath}
\usepackage{appendix}
\usepackage[ansinew]{inputenc}
\usepackage{bbm}
\usepackage{bm}
\usepackage{amsbsy}
\usepackage{dsfont} 
\usepackage{graphicx} 
\usepackage{epsfig}
\usepackage{epstopdf}
\usepackage{dsfont}
\usepackage{color}
\usepackage[colorlinks]{hyperref}
\usepackage[figure,table]{hypcap}
\usepackage{enumerate}
\hypersetup{
	bookmarksnumbered,
	pdfstartview={FitH},
	citecolor={darkgreen},
	linkcolor={darkred},
	urlcolor={darkblue},
	pdfpagemode={UseOutlines}}
\definecolor{darkgreen}{RGB}{50,190,50}
\definecolor{darkblue}{RGB}{0,0,190}
\definecolor{darkred}{RGB}{238,0,0}
\usepackage{soul}

\newcommand{\USB}{\ensuremath{U_{\hspace{-1.3pt}\protect\raisebox{0pt}{\tiny{$S\hspace*{-0.7pt}B$}}}}}
\newcommand{\SB}{\ensuremath{_{\hspace{-0.5pt}\protect\raisebox{0pt}{\tiny{$S\hspace*{-0.5pt}B$}}}}}
\newcommand{\Sys}{\ensuremath{_{\hspace{-0.5pt}\protect\raisebox{0pt}{\tiny{$S$}}}}}
\newcommand{\bath}{\ensuremath{_{\hspace{-0.5pt}\protect\raisebox{0pt}{\tiny{$B$}}}}}
\newcommand{\Sone}{\ensuremath{_{\hspace{-0.5pt}\protect\raisebox{0pt}{\tiny{$S_{1}$}}}}}
\newcommand{\Stwo}{\ensuremath{_{\hspace{-0.5pt}\protect\raisebox{0pt}{\tiny{$S_{2}$}}}}}
\newcommand{\Sonetwo}{\ensuremath{_{\hspace{-0.5pt}\protect\raisebox{0pt}{\tiny{$S_{1}S_{2}$}}}}}
\newcommand{\Soneortwo}{\ensuremath{_{\hspace{-0.5pt}\protect\raisebox{0pt}{\tiny{$S_{1}(S_{2})$}}}}}
\newcommand{\SoneorStwo}{\ensuremath{_{\hspace{-0.5pt}\protect\raisebox{0pt}{\tiny{$S_{1(2)}$}}}}}
\newcommand{\Stwoorone}{\ensuremath{_{\hspace{-0.5pt}\protect\raisebox{0pt}{\tiny{$S_{2}(S_{1})$}}}}}
\newcommand{\nl}{\ensuremath{\hspace*{-0.5pt}}}
\newcommand{\nr}{\ensuremath{\hspace*{0.5pt}}}
\newcommand{\dvline}{\ensuremath{\hspace*{0.5pt}|\hspace*{-1pt}|\hspace*{0.5pt}}}

\newcommand{\Tone}{\ensuremath{T_{\protect\raisebox{-0.5pt}{\scriptsize{$\mathrm{I}$}}}}}
\newcommand{\Ttwo}{\ensuremath{T_{\protect\raisebox{-0.5pt}{\scriptsize{$\mathrm{I\hspace*{-0.5pt}I}$}}}}}
\newcommand{\bone}{\ensuremath{\beta_{\protect\raisebox{-0.5pt}{\scriptsize{$\mathrm{I}$}}}}}
\newcommand{\btwo}{\ensuremath{\beta_{\protect\raisebox{-0.5pt}{\scriptsize{$\mathrm{I\hspace*{-0.5pt}I}$}}}}}

\newcommand{\fbra}[1]{\ensuremath{\left\langle\!\left\langle\right.\right.\! #1 \!\left.\left.\right|\hspace*{-0.75pt}\right|}}
\newcommand{\fket}[1]{\ensuremath{\left|\hspace*{-0.75pt}\left|\right.\right.\! #1 \!\left.\left.\right\rangle\!\right\rangle}}

\newcommand{\comm}[2]{\ensuremath{\left[\right.\! #1 \,, #2 \!\left.\right]}}
\newcommand{\anticomm}[2]{\ensuremath{\left\{\right.\! #1 \,, #2 \!\left.\right\}}}

\newcommand{\ie}{\textit{i.e.}}
\newcommand{\eg}{\textit{e.g.}}

\newcommand{\tr}{\textnormal{Tr}}
\newcommand{\djj}{d\kern-0.4em\char"16\kern-0.1em}

\begin{document}

\title{The thermodynamics of creating correlations: Limitations and optimal protocols}
\author{\hspace*{-2mm}David Edward Bruschi}
\thanks{D.~E.~Bruschi, M.~Perarnau-Llobet, and N.~Friis have contributed equally to this work.}
\affiliation{Racah Institute of Physics and Quantum Information Science Centre, Hebrew University of Jerusalem, 91904 Jerusalem, Israel}
\author{Mart{\'i} Perarnau-Llobet}
\thanks{D.~E.~Bruschi, M.~Perarnau-Llobet, and N.~Friis have contributed equally to this work.}
\affiliation{ICFO | The Institute of Photonic Sciences, Mediterranean Technology Park, 08860 Castelldefels (Barcelona), Spain}
\author{Nicolai~Friis}
\thanks{D.~E.~Bruschi, M.~Perarnau-Llobet, and N.~Friis have contributed equally to this work.}
\affiliation{Institute for Theoretical Physics, University of Innsbruck, Technikerstra{\ss}e 21a, A-6020 Innsbruck, Austria}
\affiliation{Institute for Quantum Optics and Quantum Information, Austrian Academy of Sciences, Technikerstra{\ss}e 21a, A-6020 Innsbruck, Austria}
\author{Karen V.~Hovhannisyan}
\affiliation{ICFO | The Institute of Photonic Sciences, Mediterranean Technology Park, 08860 Castelldefels (Barcelona), Spain}
\author{Marcus Huber}
\affiliation{Departament de F\'isica, Universitat Aut\`onoma de Barcelona,
08193 Bellaterra, Spain}
\affiliation{ICFO | The Institute of Photonic Sciences, Mediterranean Technology Park, 08860 Castelldefels (Barcelona), Spain}

\begin{abstract}
We establish a rigorous connection between fundamental resource theories at the quantum scale. Correlations and entanglement constitute indispensable resources for numerous quantum information tasks. However, their establishment comes at the cost of energy, the resource of thermodynamics, and is limited by the initial entropy. Here, the optimal conversion of energy into correlations is investigated. Assuming the presence of a thermal bath, we establish general bounds for arbitrary systems and construct a protocol saturating them. The amount of correlations, quantified by the mutual information, can increase at most linearly with the available energy, and we determine where the linear regime breaks down. We further consider the generation of genuine quantum correlations, focusing on the fundamental constituents of our universe: fermions and bosons. For fermionic modes, we find the optimal entangling protocol. For bosonic modes, we show that while Gaussian operations can be outperformed in creating entanglement, their performance is optimal for high energies.
\end{abstract}

\maketitle

\section{Introduction}

\vspace*{-2mm}
Correlations constitute fundamental resources for various tasks in quantum information processing~\cite{NielsenChuang2000}. In order to create the paradigmatic resource \textemdash entanglement \textemdash global operations are required. These operations come at a price: They require access to all of the subsystems of the target system and precise control over their interactions. This motivates the formulation of quantum information theory as~a resource theory with respect to the limitations imposed by local operations and classical communication (LOCC)~\cite{HorodeckiRPMK2007,BrandaoHorodeckiOppenheimRenesSpekkens2013,HorodeckiOppenheim2013a,EltschkaSiewert2014}.

However, there is another price to be paid for correlating quantum systems. As any amount of correlation implies extractable work~\cite{OppenheimHorodeckiMPR2002,Zurek2003,BragaRulliOliveiraSarandy2014,MaruyamaMorikoshiVedral2005,PerarnauLlobetHovhannisyanHuberSkrzypczykBrunnerAcin2014}, it follows that energy is required to establish correlations. The required energy depends on the inevitable initial entropy of the system. This establishes~a link to another resource theory \textemdash (quantum) thermodynamics, where the purity of the system, as well as the available free energy constitute fundamental resources due to the restrictions of the first and second laws of thermodynamics.

Recent interest in thermodynamics in the quantum domain (see, \eg,~\cite{MaruyamaNoriVedral2009,DelRioAbergRennerDahlstenVedral2011,HovhannisyanPerarnauLlobetHuberAcin2013,ParrondoHorowitzSagawa2015}) is, in part, fueled by this interesting connection to (quantum) information and its implications for the very foundations of thermodynamic laws \cite{BrandaoHorodeckiNgOppenheimWehner2015,LostaglioJenningsRudolph2014,WeilemannKraemerFaistRenner2015}. Combining the limitations of both theories shows that the resources of one theory are of great significance to the other as well. Examples range from an inevitable energy cost of measurements~\cite{Jacobs2012}, and the role of entanglement (and other quantum effects) in thermal machines~\cite{ScaraniZimanStelmachovicGisinBuzek2002,AlickiFannes2013,BrunnerHuberLindenPopescuSilvaSkrzypczyk2014,GallegoRieraEisert2014,CorreaPalaoAdessoAlonso2013,
CorreaPalaAlonsoAdesso2014,CorreaPalaoAdessoAlonso2014}, to scenarios~\cite{HuberPerarnauHovhannisyanSkrzypczykKloecklBrunnerAcin2014} in which thermodynamic resources play~a role in the formation of entanglement and other types of shared information.

This naturally leads us to ask two fundamental questions about the physical limitations of quantum information processing: \textit{What is the maximal amount of correlation and entanglement that can be generated for~a given energy cost?} \textit{How does the inevitable mixedness due to finite temperatures influence these costs, or, in other words, what is the role of the purity as a resource?} For closed systems, these questions were addressed in Ref.~\cite{HuberPerarnauHovhannisyanSkrzypczykKloecklBrunnerAcin2014}. Here, we extend these results by (i) considering the presence of an auxiliary thermal bath, (ii) deriving fundamental bounds and optimal protocols for the creation of total correlations, and (iii) analyzing the minimal energy cost for creating genuine quantum correlations, \ie, entanglement, in fermionic and bosonic systems.

\begin{figure}[ht!]
\label{fig:Illustration}
\includegraphics[width=0.4\textwidth]{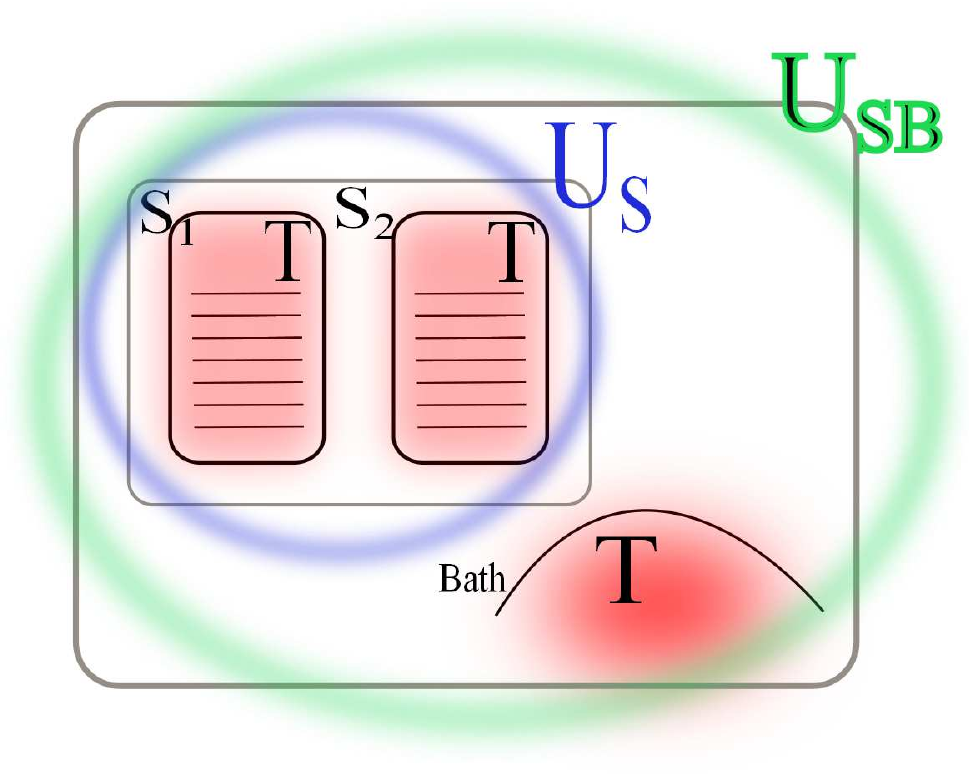}
\caption{
\textbf{Illustration of the general setup:} Two quantum systems,~$S_{1}$ and~$S_{2}$, at thermal equilibrium with~a bath at temperature~$T$ are acted upon either by~a unitary~$U\Sys$ on the bipartite system or by~a more general unitary~$\USB$ that also involves the bath. The application of these unitaries, which correlate the system, requires~a supply of external energy. In this general setting, we determine the optimal amount of correlations and entanglement that can be generated in the system for any given amount of energy.}
\end{figure}

First, assuming unlimited control over the system and an arbitrarily large thermal bath (see Fig.~\ref{fig:Illustration}), we derive the ultimate limitations for any protocol to generate correlations as quantified by the mutual information. This top\textendash down approach provides absolute bounds which cannot be outperformed, and we present~a protocol for which these bounds can be saturated.

To complement these results, we then present~a bottom\textendash up approach for the generation of entanglement between fundamental physical systems\textemdash field modes with fermionic or bosonic statistics. Taking into account limitations such as superselection rules for fermions, and using experimentally feasible and widely available techniques for bosonic modes, we provide protocols for the creation of entanglement. While we find the fermionic protocols to be optimal, we show that the practical bosonic protocols become optimal only in the limit of large input energies. Surprisingly, we find that for both the total and genuine quantum correlations, operations involving the bath may be restricted to simple thermalization processes.


\section{Framework}\label{sec:framework}
\vspace*{-2mm}
Let us start by defining some of the basic notions of quantum thermodynamics. The energy~$E$ of any quantum system~$S$ is given by the expectation value of the corresponding Hamiltonian~$H\Sys$ in the system state~$\rho$, that is, $E(\rho)=\tr(H\Sys\hspace*{0.5pt}\rho)$. A~crucial quantity, which we will refer to throughout this work, is the free energy~$F$, \ie,
\begin{align}
    F(\rho) &=\,E(\rho)\,-\,T\hspace*{0.5pt}S(\rho)\,,
    \label{eq:non equilibrium free energy}
\end{align}
where $S(\rho)=-\tr\bigl(\rho\ln(\rho)\bigr)$ is the von~Neumann entropy. The free energy defines the amount of work that is extractable from~a system when given access to a thermal bath at temperature~$T$. For thermal states~$\tau(\beta)$ of the form
\vspace*{-3mm}
\begin{align}
    \tau(\beta)    &=\,\frac{e^{-\beta H\Sys}}{\mathcal{Z}(\beta)}\,,
    \label{eq:general thermal states}
    \vspace*{-2mm}
\end{align}
the free energy takes on its minimal value $F\bigr(\tau(\beta)\bigr)=-T\,\ln(\mathcal{Z})$, where~$\mathcal{Z}$ is the partition function,~$\beta=1/T$, and we work in units where~$\hbar=k_{\protect\raisebox{-0pt}{\tiny{B}}}=1$. For arbitrary states,~$F(\rho)$ may be referred to as the nonequilibrium free energy. In the following, we consider the initial state of the system~$S$ to be thermal, $\rho\Sys=\tau\Sys(\beta)$.

We further assume that a~heat bath~$B$, that is, an arbitrarily large ancillary system in thermal equilibrium, is available. The total Hamiltonian is $H=H\Sys+H\bath\,$, and the initial state can be written as \mbox{$\tau\SB(\beta)=\tau\Sys(\beta)\otimes\hspace*{0.5pt}\tau\bath(\beta)$}. The Hilbert space~$\mathcal{H}\Sys=\mathcal{H}\Sone\otimes\mathcal{H}\Stwo$ of~$S$ is divided into two subsystems,~$S_{1}$ and $S_{2}\,$, which we assume to be noninteracting, such that~$H\Sys=H\Sone+H\Stwo$ and, consequently, $\tau\Sys(\beta)=\tau\Sone(\beta)\otimes\tau\Stwo(\beta)$. These initially uncorrelated subsystems are to be correlated via~a global unitary operation~$\USB$ on the total Hilbert space $\mathcal{H}=\mathcal{H}\Sys\otimes\mathcal{H}\bath$. The unitary~$\USB$ is the most general operation available, assuming that~$S$ and~$B$ are isolated. Any such unitary can be thought of as~a single cycle of~a quantum machine. The associated energy cost~$W$ is defined as the average overall energy change,
\vspace*{-2mm}
\begin{align}
    W   &=\,\tr\Bigl(H\bigl[\USB\hspace*{0.5pt}\tau\SB(\beta)\USB^{\dagger}-\tau\SB(\beta)\bigr]\Bigr)\,=\,\Delta E\Sys + \Delta E\bath\,,
    \vspace*{-2mm}
    \label{eq:W in}
\end{align}
and it corresponds to the total work that needs to be performed to correlate~$S$. Since~$\USB$ leaves the total entropy of~$\tau\SB$ invariant,~$W$ can be identified with the total change in free energy, which is minimal for the initial thermal state. Note that any initial state different from~a thermal state at~the temperature of the bath would provide extractable work that could be used to create correlations. To avoid this dependence on the initial state, and to properly account for the work invested in the system, we chose an initial thermal state at temperature~$T$, corresponding to the temperature of the heat bath. It follows that $W \geq 0$, and hence any operation~$\USB$ requires some energy. The aim of this paper is to determine how this energy may be used most efficiently to correlate the systems~$S_{1}$ and~$S_{2}\,$.

We distinguish two kinds of correlations: total correlations, and genuine quantum correlations (entanglement). We quantify the former by the mutual information
\begin{align}
    \mathcal{I}\Sonetwo(\rho\Sys) &=\,S(\rho\Sone)\,+\,S(\rho\Stwo)\,-\,S(\rho\Sys)\,,
    \label{eq:mutual inf definition}
\end{align}
which measures the amount of global information shared among the systems~$S_{1}$ and~$S_{2}$, \ie, the information encoded within the state~$\rho\Sys$ that is not accessible through its subsystems alone. Pure quantum states for which the mutual information is nonzero are entangled, but this is not necessarily the case for mixed states. To quantify genuine quantum correlations between~$S_{1}$ and~$S_{2}$, we employ the entanglement of formation (see, \eg, Ref.~\cite{PlenioVirmani2007} for~a review of available entanglement measures), which can be defined as the minimal average mutual information across all decompositions of the mixed quantum state into pure state ensembles, \ie,
\begin{align}
    E_{oF}(\rho\Sys)  &:=\frac{1}{2}\inf_{\mathcal{D}(\rho\Sys)}\sum\limits_{i} p_{i}\,\, \mathcal{I}\Sonetwo\hspace*{-0.5pt}\bigl(\ket{\psi_{i}}\!\bra{\psi_{i}}\bigr)\,,
    \label{eq:entanglement of formation definition}
\end{align}
where $\mathcal{D}(\rho\Sys)=\left\{p_{i},\ket{\psi_{i}}|\sum_{i}p_{i}\ket{\psi_{i}}\!\bra{\psi_{i}}=\rho\Sys\right\}$. In~a finite-dimensional system, the entanglement of formation represents the number of maximally entangled states per copy that are needed asymptotically to create the state via LOCC.


\section{Correlating quantum systems: energy cost and optimal protocols}\label{subsec:Correlating quantum systems: energy cost and optimal protocols}

We now present our main results. We start with the top\textendash down approach, where we determine the ultimate limitations of creating correlations, as quantified by the mutual information. Using the facts that the initial thermal state is completely uncorrelated, $S(\tau\Sys)=S(\tau\Sone)+S(\tau\Stwo)$, and that the global unitary leaves the overall entropy invariant, $S(\USB\tau\SB\USB^{\dagger})=S(\tau\SB)$, we combine Eqs.~(\ref{eq:non equilibrium free energy}) and~(\ref{eq:W in}) to express the energy cost~$W$ in terms of the free energy difference as
\begin{align}
    W   &=\,\Delta F\Sys\,+\,\Delta F\bath\,+\,T\,\mathcal{I}\SB\,,
    \label{eq:free energy difference S vs B}
\end{align}
obtaining~a similar expression to those discussed, \eg, in Refs.~\cite{OppenheimHorodeckiMPR2002,OppenheimHorodeckiKMPR2003,EspositoVanDenBroeck2011,ReebWolf2014} in related contexts. A~detailed derivation of Eq.~(\ref{eq:free energy difference S vs B}) can be found in Appendix~\ref{sec:methods B Energy cost of a general unitary}. In complete analogy to~(\ref{eq:free energy difference S vs B}), we may split~$\Delta F\Sys$ into the free energy differences of its subsystems, and their correlation as
\begin{align}
    \Delta F\Sys    &=\,\Delta F\Sone\,+\,\Delta F\Stwo\,+\,T\,\mathcal{I}\Sonetwo\,,
    \label{eq:free energy difference S1 vs S2}
\end{align}
for which~a proof is also given in Appendix~\ref{sec:methods B Energy cost of a general unitary}. For any thermal state~$\tau$, the free energy difference to another (non-equilibrium) state~$\rho$ may be expressed through the relative entropy~$S(\rho\dvline\tau)=-S(\rho)-\tr(\rho\ln\tau)$ as~$\Delta F=T\,S\bigl(\rho\dvline\tau(\beta)\bigr)$. This, in turn, allows us to write~$W$ in the form
\begin{align}
    \beta\,W   &=\,S(\rho\Sone\dvline\tau\Sone)+S(\rho\Stwo\dvline\tau\Stwo)+S(\rho\bath\dvline\tau\bath)\nonumber\\[1mm]
    &\ +\,\mathcal{I}\Sonetwo\,+\,\mathcal{I}\SB\,,
    \label{eq:W in terms of entropies}
\end{align}
where~$\rho\Sone$,~$\rho\Stwo$, and~$\rho\bath$ denote the final reduced states for the subsystems,~$S_{1}$ and~$S_{2}$, and the bath~$B$, respectively. In other words, work can be invested to shift the thermal marginals away from equilibrium or to create correlations. Since all quantities on the right-hand side of Eq.~(\ref{eq:W in terms of entropies}) are non-negative, it can be immediately inferred that the following ultimate bound holds for the amount of correlation that can be generated between the subsystems for~a given energy cost~$W$ and temperature~$T=1/\beta$:
\begin{align}
    \mathcal{I}\Sonetwo   &\leq\, \beta\,W\,.
    \label{eq:ultimate bound}
\end{align}

\begin{figure}[ht!]
\label{fig:Protocol}
\includegraphics[width=0.45\textwidth]{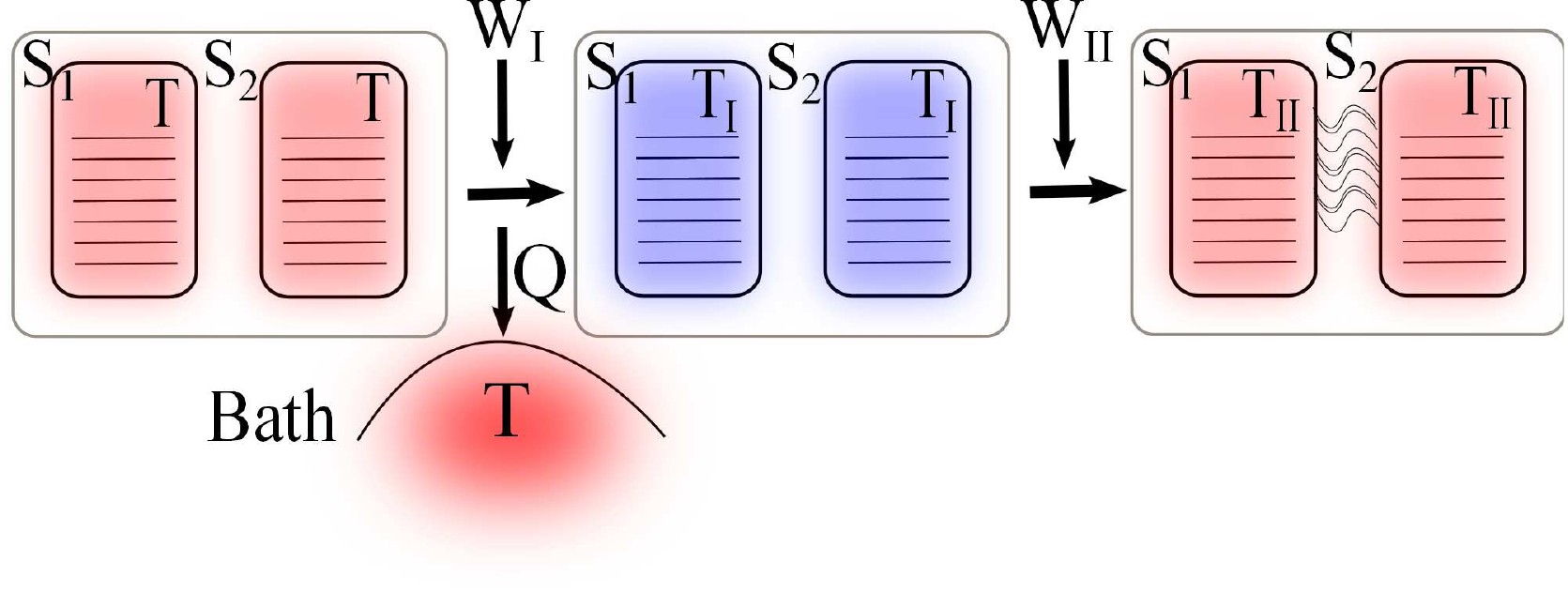}
\caption{
\textbf{Illustration of the protocol:} In the first step the system is cooled down by~a controlled interaction with the bath, and the heat~$Q$ is transferred to the bath. The associated work cost is $W_\mathrm{I}$. In the second step, the system is isolated from the bath before it is correlated though~a unitary operation, which effectively heats up the subsystems. The energy cost of the second step is $W_{\mathrm{I\nl I}}$.}
\end{figure}

Remarkably, it is possible to saturate this bound using~a simple set of operations: unitary operations on~$S$ and interactions with the bath to thermalize the system. These operations are enough to obtain $W=\Delta F\Sys$ in (\ref{eq:free energy difference S vs B}) in the limit of an arbitrarily large bath that is complex enough to thermalize
the system each time they come in contact (see Ref.~\cite{Aberg2013} for~a proof, and Ref.~\cite{SkrzypczykShortPopescu2014} for~a description in terms of unitary operations). We are now ready to present the protocol achieving $W=T\nr\mathcal{I}\Sonetwo$, which can be divided into two steps (see Fig.~\ref{fig:Protocol}).\\

\newpage

\begin{enumerate}[{(I)}]
\item{\textbf{Cooling}:\
    First, the temperature of~$S$ is lowered from~$T$ to~$\Tone\leq T$, reducing the global entropy of the system. The (minimal) energy cost for this thermalization process is $ W_{\mathrm{I}}=\Delta F\Sys$, \ie,
    \begin{align}
        W_{\mathrm{I}} &=\,F\bigl(\tau\Sys(\bone)\bigr)-F\bigl(\tau\Sys(\beta)\bigr)\,,
        \label{eq:work step I}
    \end{align}
    where $\bone=1/\Tone$.}
\end{enumerate}
\begin{enumerate}[{\hspace*{-5mm}(I\hspace*{-0.5pt}I)}]
\item{\textbf{Correlating}:\
    In the second step, the system is isolated from the bath and it is correlated via a unitary operation $U_{\mathrm{corr}}$. Following Ref.~\cite{HuberPerarnauHovhannisyanSkrzypczykKloecklBrunnerAcin2014}, the unitary is chosen such that~$S_{1}$ and~$S_{2}$ are locally thermal at temperature~$\Ttwo=1/\btwo\geq\Tone$, \ie,
    \vspace*{-1mm}
    \begin{align}
        \hspace*{6mm}\tr\Soneortwo\bigl(U_{\mathrm{corr}}\nr\tau\Sys(\bone)\nr U_{\mathrm{corr}}^{\dagger}\bigr) &=\,\tau\Stwoorone(\btwo)\,.
        \label{eq:thermal marginals step II}
    \end{align}
    This choice ensures that the systems are correlated at minimal energy cost~$W_{\mathrm{I\nl I}}$, see~\cite{HuberPerarnauHovhannisyanSkrzypczykKloecklBrunnerAcin2014}.}
\end{enumerate}

There is thus~a tradeoff between the amount of work~$W_{\mathrm{I}}$, invested to cool down the system, which allows one to potentially obtain larger correlations, and the work~$W_{\mathrm{I\nl I}}$, invested to actually correlate it. As we show in detail in Appendix~\ref{sec:methods C Optimal protocol for generating mutual information}, both contributions add up to
\begin{align}
    W   &=\,W_{\mathrm{I}}\,+\,W_{\mathrm{I\nl I}}\,=\,
    T\,\mathcal{I}\Sonetwo\,+\,T\,S\bigl(\tau\Sys(\btwo)\dvline\tau\Sys(\beta)\bigr)\,.
    \label{eq:optimal correlation protocol work}
\end{align}
Therefore, optimality is achieved when the local temperature of the final state marginals is identical to the initial temperature, $\Ttwo=T$, such that $W=T\nr\mathcal{I}\Sonetwo$.

However, it may occur that this condition would require more energy to be used in the first step than is needed to reach the ground state. In such~a case, the excess energy can be put to better use further correlating the final state, raising the local temperatures of the subsystems beyond~$\Ttwo=T$. These considerations yield a more precise bound (see Appendix~\ref{sec:methods C Optimal protocol for generating mutual information}), given by
\begin{align}
\mathcal{I}\Sonetwo &\leq
    \begin{cases}
        \beta\nr W     &\ \mbox{if}\ \beta\nr W \leq  S\bigl(\tau\Sys(\beta)\bigr)\,,  \\
        S(\tau\Sys(\btwo)) &\ \mbox{if}\ \beta\nr W >  S\bigl(\tau\Sys(\beta)\bigr)\,,
    \end{cases}
    \label{eq:energy regimes mutual inf}
\end{align}
where~$\btwo$ is given by the implicit relation $E\bigl(\tau\Sys(\btwo)\bigr)=W+F\bigl(\tau\Sys(\beta)\bigr)$. There are hence two distinct regimes. When an energy smaller than $TS\bigl(\tau\Sys(\beta)\bigr)$ is supplied, the correlations scale linearly with the work input. As more energy is provided, additional work needs to be invested to move the states further out of local equilibrium, leading to noticeably different behavior. For instance, for two bosonic modes, the correlations scale logarithmically with the work input for $\beta W \gg  S\bigl(\tau\Sys(\beta)\bigr)$, as we show in Appendix~\ref{sec:methods E Generation of mutual information in two bosonic modes}.

Finally, it is worth mentioning that our protocol is extendible to nonequilibrium initial states. One then needs to first extract the work content of the state, which leaves it in~a thermal state at the temperature of the bath. Our protocol can then be readily applied using the extracted work in addition to any externally supplied energy to correlate the system.


\section{Energy cost of entanglement generation}\label{sec:energy cost of entanglement generation}

Having provided general bounds on the energy cost of correlating two arbitrary systems, we now turn to the case of genuine quantum correlations, \ie, entanglement. Here the situation is much more complex. Even determining whether~a given quantum state is separable or not is generally NP hard. Therefore, obtaining a~general solution for arbitrary systems is~a daunting task that seems intractable. We therefore complement the previous top-down approach for general correlations by pursuing~a bottom-up strategy to investigate the energy cost for generating entanglement. We focus our attention on two physically relevant cases, namely, systems of two fermionic or bosonic modes. For the low-dimensional fermionic problem and the case of bosonic Gaussian states, computing the entanglement of formation in Eq.~(\ref{eq:entanglement of formation definition}) becomes feasible.

Besides making the problem more tractable, the very interesting features of bosonic and fermionic systems further motivate our choice. On one hand, modes of quantum fields play~a fundamental role in the description of nature in the context of (relativistic) quantum theory. Hence, they provide~a more general framework for our analysis than systems with~a fixed number of particles, which appear as secondary quantities, \ie, as excitations of the modes in question. On the other hand, this approach allows us to analyze the role of fermionic and bosonic particle statistics, and the corresponding finite and infinite-dimensional Hilbert spaces for two modes. In addition, the formulation in terms of individual mode operators naturally lends itself to the Hamiltonian structure, giving~a clear interpretation to the involved energy costs.

In this section, we consider protocols along the same lines as previously, \ie, first varying the temperature of the systems (not necessarily symmetrically) and then correlating them via unitary operations. This choice is well justified because any other operation that would either create correlations between the system and the bath or significantly change the state of the bath would have~a higher energy cost, as can be seen from Eq.~(\ref{eq:free energy difference S vs B}).


\subsection{Fermionic systems}\label{sec:fermionic systems main}

We now consider~a finite-dimensional system, two modes of (equal) frequency~$\omega$ of an uncharged, noninteracting fermionic field. On one hand, the simplicity of this system allows us to determine the amount of entanglement that may be generated for any given amount of energy. On the other hand, several conceptually interesting features arise from the fermionic algebra, that is, the mode operators~$b_{\raisebox{-1.5pt}{\tiny{1}}}$,~$b_{\raisebox{-0.5pt}{\tiny{1}}}^{\dagger}$,~$b_{\raisebox{-1.5pt}{\tiny{2}}}$, and~$b_{\raisebox{-0.5pt}{\tiny{2}}}^{\dagger}$ satisfy the anticommutation relations $\anticomm{b_{m}}{b_{n}^{\dagger}}=\delta_{mn}$ and $\anticomm{b_{m}}{b_{n}}=0$, where $m,n=1,2\,$. The Hamiltonian of the system is (up to~a constant) given by $H\Sys=H\Sone+H\Stwo=\omega\bigl(b_{\raisebox{-0.5pt}{\tiny{1}}}^{\dagger}b_{\raisebox{-1.5pt}{\tiny{1}}}+b_{\raisebox{-0.5pt}{\tiny{2}}}^{\dagger}b_{\raisebox{-1.5pt}{\tiny{2}}}\bigr)$. To distinguish the fermionic and bosonic case, we denote the fermionic Fock states by double-lined kets, \eg, the vacuum state is written as~$\fket{\!0\!}$. The single-particle states are obtained by the action of the creation operators, \ie, $\fket{\!1_{m}\!}=b_{m}^{\dagger}\fket{\!0\!}$. We define the two-particle state via
$\fket{\!1\Sone\!}\fket{\!1\Stwo\!}=b_{\raisebox{-0.5pt}{\tiny{1}}}^{\dagger}b_{\raisebox{-0.5pt}{\tiny{2}}}^{\dagger}\fket{\!0\!}$, where we have omitted the symbol for the antisymmetrized tensor product on the left-hand side (see Refs.~\cite{FriisLeeBruschi2013} or~\cite[pp.~37]{Friis:PhD2013} for more details on the notation used here and the fermionic Fock space). The system we investigate here obeys Fermi-Dirac statistics, and the partition function is hence $\mathcal{Z}_{\mathrm{FD}}(\beta)=\bigl(1+e^{-\beta}\bigr)$, and we specify temperatures in units of~$\omega$ [recall, that~$(\hbar=k_{\protect\raisebox{-0pt}{\tiny{B}}}=1$] from now on. The average initial particle numbers are given by~$N\Soneortwo=\tr\bigl(b_{\raisebox{-0.5pt}{\tiny{1(2)}}}^{\dagger}b_{\raisebox{-0.5pt}{\tiny{1(2)}}}\tau\Sys\bigr)$. The fermionic two-mode thermal state may then be expressed as
\begin{align}
    \tau\Sys    &=\frac{e^{-\beta}}{\mathcal{Z}_{\mathrm{FD}}^{2}}\Bigl(
        e^{\beta}\fket{\!0\!}\!\fbra{\!0\!}+
        \fket{\!1\Sone\!}\!\fbra{\!1\Sone\nl\!}+\fket{\!1\Stwo\!}\!\fbra{\!1\Stwo\!}
        \nonumber\\[1mm]
    &\ \ +\,e^{-\beta}\fket{\!1\Sone\!}\!\fket{\!1\Stwo\!}\!\fbra{\!1\Stwo\!}\!\fbra{\!1\Sone\hspace*{-1pt}\!}\Bigr)\,.
        \label{eq:fermion two-mode thermal state}
\end{align}

With these preliminaries at hand, we consider protocols along the lines of that presented in Section~\ref{subsec:Correlating quantum systems: energy cost and optimal protocols} to create entanglement. In the first step of such~a procedure, using the interaction with the bath, the temperature of the two modes is lowered as before, which manifests in altered particle numbers~$N^{\mathrm{I}}\Sone$ and~$N^{\mathrm{I}}\Stwo$. The energy cost~$W_{\mathrm{I}}$ for this step is given by the free energy difference to the transformed state.

\begin{figure}[ht!]
(a)\includegraphics[width=0.47\textwidth]{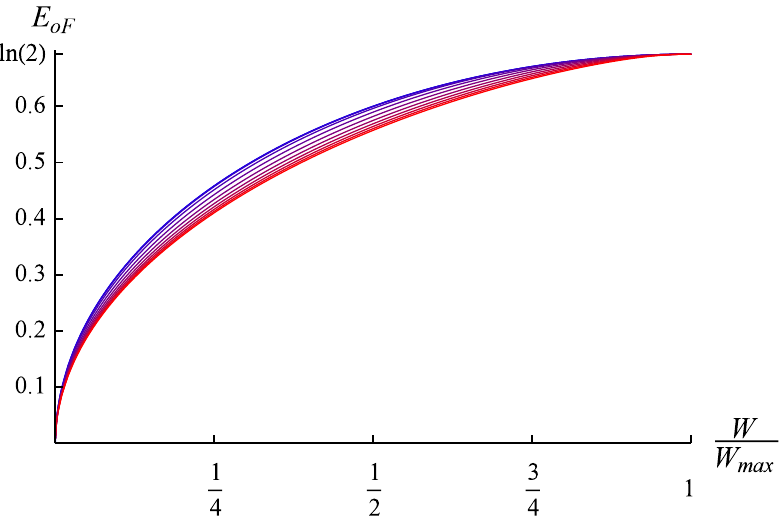}
(b)\includegraphics[width=0.47\textwidth]{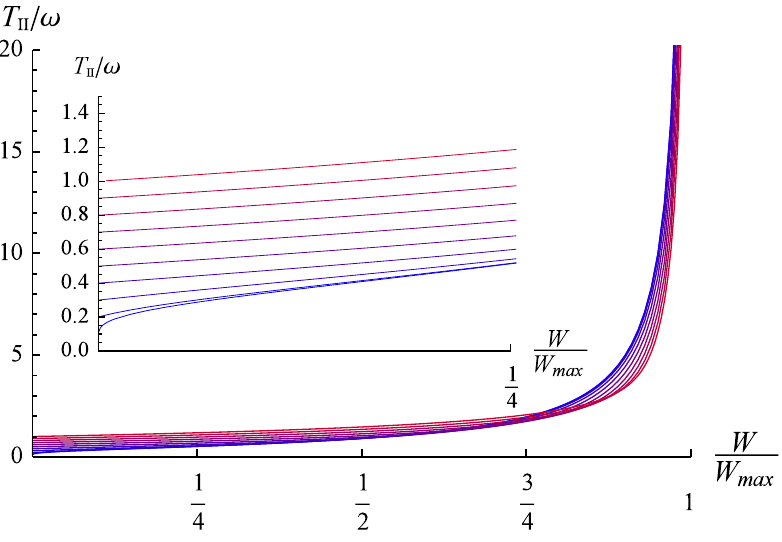}
\caption{\label{fig:fermions optimal even entanglement with bath}
\textbf{Fermionic entanglement cost}: The solid curves in Fig.~\ref{fig:fermions optimal even entanglement with bath}~(a) show the amount of entanglement (of formation) that can maximally be generated in the even subspace of two fermionic modes that are initially in~a thermal state, for~a given energy cost~$W$. The curves are plotted for initial temperatures varying from $T=0$ to $T=1$ in steps of $0.1$ (top to bottom) in units~$\hbar\omega/k_{\protect\raisebox{-0pt}{\tiny{B}}}$. The horizontal axis shows the relative energy cost, \ie, the fraction of~$W$ and the minimal energy cost~$W_{\mathrm{max}}=\bigl[2T\ln\bigl(e^{\beta}+1\bigr)-\omega\bigr]$ to generate~a maximally entangled pure state. Fig.~\ref{fig:fermions optimal even entanglement with bath}~(b) shows the corresponding effective final temperature~$\Ttwo\geq T$ of the marginals after the protocol.}
\end{figure}

In the second step of the protocol, unitaries on the two-mode space~$S$ are applied to correlate the system. In the case of fermionic modes, these operations are further restricted by superselection rules. Since the state of any single fermion acquires~a phase of~$\pi$ upon~a rotation around~$2\pi$, rotational symmetry prohibits coherent superpositions of even and odd numbers of fermions. Moreover, the superselection rules modify the definition of the entanglement of formation of Eq.~(\ref{eq:entanglement of formation definition}) in the sense that the minimization is carried out only over pure state ensembles that respect superselection~\cite{CabanPodlaskiRembielinskiSmolinskiWalczak2005}. We hence take as~a measure of entanglement the minimum number, per copy, of maximally entangled states of the two fermionic modes, which are needed to assemble~a given two-mode state. As is shown in Appendix~\ref{sec:Optimal protocol for fermionic entanglement of formation}, this well-defined measure of entanglement can be expressed by the energy cost~$W_{\mathrm{I\nl I}}$ of the correlating step as
\begin{align}
    E_{oF} &=\,\ln(2)\,
    \sqrt{\frac{W_{\mathrm{I\hspace*{-0.5pt}I}}}{\omega}}
    \sqrt{2\frac{e^{\beta_{\mathrm{I}}}-1}{e^{\beta_{\mathrm{I}}}+1}-\frac{W_{\mathrm{I\hspace*{-0.5pt}I}}}{\omega}}\,.
    \label{eq:EoF fermions even protocol}
\end{align}
Similar to the previous section, we determine the optimal splitting of~$W$ into~$W_\mathrm{I}$ and~$W_{\mathrm{I\nl I}}$, and we express it in terms of the optimal final temperature~$\Ttwo$.  The results of this numerical optimization are presented in Fig.~\ref{fig:fermions optimal even entanglement with bath}. Although the protocol is very similar to that for the generation of mutual information, optimality is not achieved for $\Ttwo=T$, but rather when~$\Ttwo\geq T$, see Fig.~\ref{fig:fermions optimal even entanglement with bath}~(b).

One can further improve upon these results by taking advantage of the peculiar properties of fermionic entanglement, in particular the existence of mixed, maximally entangled states~\cite{DArianoManessiPerinottiTosini2014b}. These particularities may occur because the subspaces of even and odd fermion numbers decouple. Consequently, no unitaries may introduce correlations between these subspaces. The optimally correlating unitary $U_{\mathrm{corr}}$ can therefore be decomposed into two independent rotations. Furthermore, we find that altering the temperatures of the subsystems asymmetrically, \ie, cooling one mode while heating the other, can be beneficial. Allowing for such asymmetric temperatures, we numerically optimize the fermionic entanglement of formation generated at~a fixed energy cost. The results are discussed in detail in Appendix~\ref{sec:Optimal protocol for fermionic entanglement of formation}.

\subsection{Bosonic systems}\label{sec:bosonic systems main}

Let us now investigate the optimal generation of entanglement for~a bosonic system. Analogously to the fermionic case, we consider two modes of an uncharged, noninteracting bosonic field. We assume that these modes, again labeled~$S_{1}$ and~$S_{2}$, have the same frequency~$\omega$. The corresponding annihilation and creation operators~$a_{\raisebox{-1.5pt}{\tiny{1}}}$,~$a_{\raisebox{-0.5pt}{\tiny{1}}}^{\dagger}$,~$a_{\raisebox{-1.5pt}{\tiny{2}}}$, and~$a_{\raisebox{-0.5pt}{\tiny{2}}}^{\dagger}$ satisfy the commutation relations $\comm{a_{m}}{a_{n}^{\dagger}}=\delta_{mn}$ and $\comm{a_{m}}{a_{n}}=0$, where $m,n=1,2\,$. The system Hamiltonian may be written in terms of these operators (up to~a constant) as $H\Sys=H\Sone+H\Stwo=\omega\bigl(a_{\raisebox{-0.5pt}{\tiny{1}}}^{\dagger}a_{\raisebox{-1.5pt}{\tiny{1}}}+a_{\raisebox{-0.5pt}{\tiny{2}}}^{\dagger}a_{\raisebox{-1.5pt}{\tiny{2}}}\bigr)$. The infinite-dimensional Fock space of these two modes is spanned by the vacuum state~$\ket{0}$, which is annihilated by~$a_{\raisebox{-1.5pt}{\tiny{1}}}$ and~$a_{\raisebox{-1.5pt}{\tiny{2}}}$, and the particle states, which are obtained by applying the creation operators~$a_{\raisebox{-0.5pt}{\tiny{1}}}^{\dagger}$ and~$a_{\raisebox{-0.5pt}{\tiny{2}}}^{\dagger}$ to the vacuum. The bosonic excitations obey Bose-Einstein statistics, where the partition function is given by $\mathcal{Z}_{\mathrm{BE}}(\beta)=\bigl(1-e^{-\beta}\bigr)^{-1}$. Note that the temperatures are again given in units of~$\omega$ and we have set~$\hbar=k_{\protect\raisebox{-0pt}{\tiny{B}}}=1$.

To handle this infinite-dimensional system, we will restrict our analysis of entanglement generation to Gaussian states, which commonly feature in applications in quantum information~\cite{AdessoRagyLee2014} and quantum computing~\cite{Weedbrooketal2012}, to name but~a few.
The correlations of two-mode Gaussian states can be completely described by~a real, $4\times4$ covariance matrix~$\sigma\Sys$. This matrix collects the expectation values of quadratic combinations of the mode operators\textemdash the second moments\textemdash and we may assume that the expectation values of all linear combinations of mode operators\textemdash the first moments\textemdash vanish. For~a given state~$\rho\Sys$, the components of~$\sigma\Sys$ are $(\sigma\Sys)_{mn}=\tr\bigl(\anticomm{\mathbb{X}_{m}}{\mathbb{X}_{n}}\rho\Sys\bigr)$, with the quadrature operators~$\mathbb{X}_{(2n-1)}=(a_{n}+a_{n}^{\dagger})/\sqrt{2}$ and~$\mathbb{X}_{(2n)}=-i(a_{n}-a_{n}^{\dagger})/\sqrt{2}$, and $m,n=1,2$. For the initial thermal state at temperature~$T$ that we consider here, the covariance matrix is proportional to the identity operator, $\sigma\Sys=\nu(T)\,\mathds{1}_{4}$, where the symplectic eigenvalue~$\nu$ is given by~$\nu(T)=\coth(\beta/2)$.

In the first step of the protocol to optimally generate entanglement, the initial temperature is lowered from~$T$ to ~$\Tone<T$, after which the state is represented by~$\sigma\Sys^{\mathrm{I}}=\nu^{\mathrm{I}}\,\mathds{1}_{4}$, where~$\nu^{\mathrm{I}}=\nu(\Tone)$. The energy cost for this step is given by
\begin{align}
        \frac{W_{\mathrm{I}}}{\omega}   &=\,\nu^{\mathrm{I}}-\nu(T)-2\beta^{-1}\Bigl[f\bigl(\nu^{\mathrm{I}}\bigr)-f\bigl(\nu(T)\bigr)\Bigr]\,,
        \label{eq:step I energy cost bosons}
\end{align}
where the entropy of~a two-mode thermal state represented by~$\sigma$ is expressed as $S(\sigma)=2f(\nu)=(\nu+1)\ln\bigl(\tfrac{\nu+1}{2}\bigr)-(\nu-1)\ln\bigl(\tfrac{\nu-1}{2}\bigr)$.

\begin{figure}
(a)\includegraphics[width=0.47\textwidth]{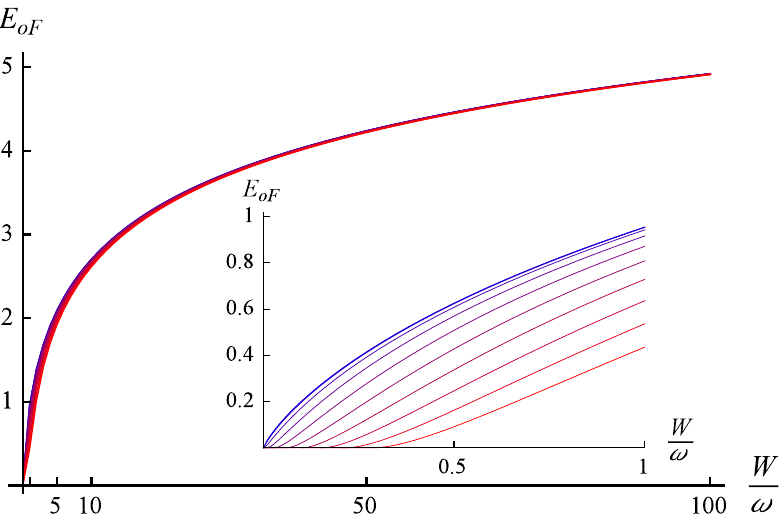}
(b)\includegraphics[width=0.47\textwidth]{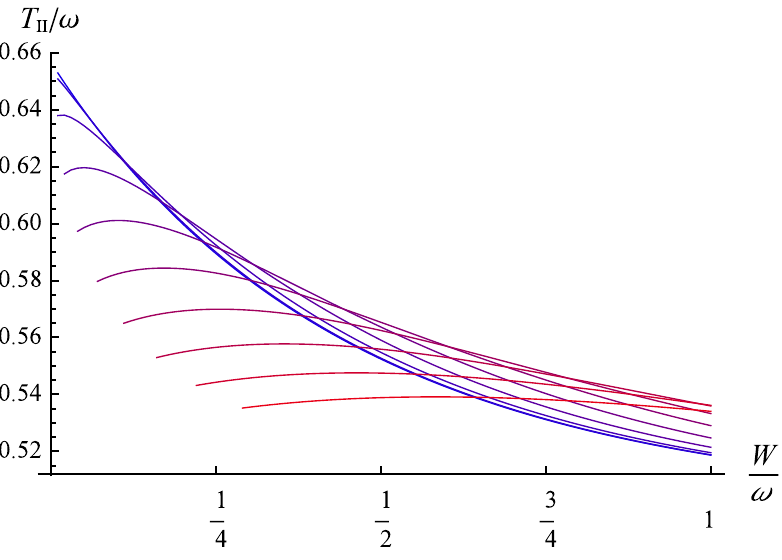}
\caption{\label{fig:bosons optimal EOF with bath}
\textbf{Optimal bosonic entanglement}: The curves in Fig.~\ref{fig:bosons optimal EOF with bath}~(a) show the optimal amount of entanglement (of formation) that can be generated by Gaussian operations on two bosonic modes,~$\Sone$ and~$\Stwo$, of frequency~$\omega$, which are initially in~a thermal state of temperature~$T$. The horizontal axis shows the supplied energy~$W$ in units of~$\omega$. Fig.~\ref{fig:bosons optimal EOF with bath}~(b) shows the local temperature~$T_{\mathrm{I\nl I}}$ of the modes after the protocol for values of~$W$ for which entanglement can be generated. The curves in both Fig.~\ref{fig:bosons optimal EOF with bath}~(a) and~(b) are plotted for initial temperatures varying from $T=0$ to $T=1$ in steps of $0.1$ (top to bottom) in units~$\hbar\omega/k_{\protect\raisebox{-0pt}{\tiny{B}}}$.}
\end{figure}

In the second step of the protocol, we restrict the entangling unitaries to Gaussian operations, which may be represented as linear transformations of the mode operators. Since the initial covariance matrix is proportional to that of the vacuum, the final covariance matrix must be proportional to that of~a pure, two-mode Gaussian state, which is locally equivalent to~a two-mode squeezed state. We may therefore conclude that the optimal Gaussian entangling operations for this situation are two-mode squeezing transformations. Moreover, throughout the protocol, the state remains symmetric with respect to the two subsystems, that is, their entropies are identical. For such states, all entanglement measures depend on~a single parameter~$\tilde{\nu}_{-}$, the smallest symplectic eigenvalue of the partial transpose. In terms of~$\tilde{\nu}_{-}$, the entanglement of formation takes the form
\begin{align}
    E_{oF}  &=\,
    \begin{cases}
    \mathfrak{h}(\tilde{\nu}_{-})\,,  &   \ \ \mbox{if}\ \ 0\,\leq\,\tilde{\nu}_{-}\,<\,1\,,\\[0.5mm]
    0\,,   &   \ \ \mbox{if}\ \ \tilde{\nu}_{-}\geq\,1\,,
    \end{cases}
    \label{eq:EoF bosons}
\end{align}
where $\mathfrak{h}(x)=h_{+}(x)\ln\bigl(h_{+}(x)\bigr)-h_{-}(x)\ln\bigl(h_{-}(x)\bigr)$, and $h_{\pm}(x)=\frac{(x\pm1)^{2}}{4x}$. One may also relate~$\tilde{\nu}_{-}$ to the squeezing parameter~$r$ of the thermal two-mode squeezed state after step~$\mathrm{I\nl I}$ via~$e^{-2r}=\tilde{\nu}_{-}/\nu^{\mathrm{I}}$, while the final state energy is given by~$\omega\bigl(\nu^{\mathrm{I}}\cosh(2r)-1\bigr)$. With this, the energy cost for step~$\mathrm{I\nl I}$ can be expressed as
\begin{align}
    \frac{W_{\mathrm{I\nl I}}}{\omega}  &=\,\frac{(\nu^{\mathrm{I}})^{2}}{2\tilde{\nu}_{-}}\Bigl[\frac{\tilde{\nu}_{-}}{\nu^{\mathrm{I}}}-1\Bigr]^{2}\,.
    \label{eq:step II work bosons}
\end{align}
Conversely, Eq.~(\ref{eq:step II work bosons}) allows us to express~$\tilde{\nu}_{-}$, and hence~$E_{oF}$, in terms of~$\nu^{\mathrm{I}}$ and~$W_{\mathrm{I\nl I}}=W-W_{\mathrm{I}}$. The results of the numerical optimization of the entanglement of formation over~$\nu^{\mathrm{I}}$ are shown in Fig.~\ref{fig:bosons optimal EOF with bath}. Note that in contrast to the fermionic case, here we find $\Ttwo<T$. Another interesting feature of the bosonic system is that for nonzero initial temperatures, entanglement cannot be generated for arbitrarily small amounts of supplied energy~\cite{BruschiFriisFuentesWeinfurtner2013}. Instead, entanglement is only created when the constraint~$(\nu^{\mathrm{I}}-1)^{2}<2W_{\mathrm{I\nl I}}/\omega$ is satisfied.

Finally,~a comment about the optimality of Gaussian operations is in order. As we show in detail in Appendix~\ref{sec:methods E About the optimality of gaussian operations}, there are two energy regimes. In the low-energy regime, Gaussian operations may be outperformed by non-Gaussian operations in generating entanglement. We provide~a protocol which achieves this, and allows leaving the separable states even for arbitrarily small amounts of supplied energy. In the high-energy regime, on the other hand, Gaussian operations are shown to be asymptotically optimal for the generation of entanglement. This can be understood in the following way. When enough energy is supplied, the ground state is reached in the cooling phase. All remaining energy can then be optimally used for Gaussian entangling operations. When large amounts of energy are invested, the fraction of the energy that is suboptimally spent in the cooling stage becomes negligible, vanishing in the limit of an infinite energy supply.


\section{Conclusion}\label{sec:discussion}

We have investigated the equivalence between free energy and the ability to create correlations in quantum systems. Any amount of correlation implies that extractable work is present in the system. Conversely, the creation of any amount of correlation comes at the price of investing work. Following this premise, we have introduced protocols that are optimal for the generation of correlations, as well as genuine quantum correlations, at minimal energy cost. For total correlations, as quantified by the mutual information, we have presented~a protocol that is optimal for arbitrary bipartite systems.

For the case of genuine quantum correlations \textemdash\ entanglement, the paradigmatic quantum resource \textemdash\ we have focused on two fermionic or two bosonic modes. For both types of systems, we have derived optimal protocols for the generation of entanglement as quantified by the well-known entanglement of formation. In the case of bosons, we have restricted the optimization to the set of Gaussian operations for the sake of feasibility. To place this choice in an appropriate context, we have also discussed explicit protocols that make use of non-Gaussian operations, showing that they can decrease the energy cost when the available energy is small. Nonetheless, our findings further show that Gaussian operations become optimal in the limit of large available energies. A~common feature of all the mentioned protocols is their remarkably simple structure. They make use of the interaction with~a thermal bath to cool (or heat) the (sub)system, which, interestingly, requires only elementary thermalization processes, before introducing correlations.

Our results connect two important resource theories, revealing the implicit thermodynamical cost and value of quantum correlations. While we have focused our efforts on
bipartite quantum systems, the results concerning correlations have the potential for~a straightforward generalization to the multipartite case when considering correlations quantified by $S(\rho)-\sum_{i} S(\rho_{i})$ where $\rho_{i}=\tr_{j\neq i}(\rho)$. Such considerations are possible extensions of our work, especially when connected to cases of multipartite entanglement generation. Here, the focus on bipartite entanglement has guaranteed the utility of the created resources for quantum communication, whereas future work concerning multipartite entanglement should be approached with great care, as generic generation of entanglement may be less useful than previously believed~\cite{GrossFlammiaEisert2009}. Other possible directions inspired by our work include similar considerations for
single-shot scenarios as, \eg, in Refs.~\cite{BrandaoHorodeckiNgOppenheimWehner2015,HorodeckiOppenheim2013b,Aberg2013}, which effectively means focusing on different entropies in the mutual information, the inclusion of catalytic systems \cite{BrandaoHorodeckiNgOppenheimWehner2015}, or even the interesting connection with the thermodynamic properties of transformations induced by nonuniform motion~\cite{FriisBruschiLoukoFuentes2012,BruschiDraganLeeFuentesLouko2013}.

\begin{acknowledgments}
We thank Gerardo Adesso, Jakob Bekenstein, John Calsamiglia, Andrzej Dragan, Markus Heyl, Nana Liu,  Vedran Dunjko, Paul Skrzypczyk, Zolt{\'a}n Zimbor{\'a}s and the LIQUID collaboration for fruitful comments and discussions. D.~E.~B. thanks ICFO and Universitat Aut\`{o}noma de Barcelona for hospitality. D.~E.~B. was supported by the I-CORE Program of the Planning and Budgeting Committee and the Israel Science Foundation (Grant No.~1937/12), as well as by the Israel Science Foundation personal grant No.~24/12. M.~P.-L. acknowledges funding from the Severo Ochoa program and the Spanish Grant No.~FPU13/05988. N.~F.~acknowledges funding by the Austrian Science Fund (FWF) through the SFB FoQuS: F4012. M.~H. acknowledges funding from the Marie Curie Grant No.~N30202 "Quacocos", from the Spanish MINECO through Project No.~FIS2013-40627-P and the Juan de la Cierva fellowship (JCI 2012-14155), from the Generalitat de Catalunya CIRIT Project No.~2014 SGR 966, and from the EU STREP-Project “RAQUEL". M.~P.-L., K.~V.~H. and M.~H. are grateful for support from the EU COST Action No.~MP1209, "Thermodynamics in the quantum regime".

\end{acknowledgments}


\appendix
\section*{Appendix}
\renewcommand\appendixname{}


\section{Preliminaries}\label{sec:methods A preliminaries}

Before we present detailed proofs for the main results, let us review some preliminary concepts. First, recall that the free energy of~a state~$\rho$ is given by
\begin{align}
    F(\rho) &=\,E(\rho)\,-\,T\hspace*{0.5pt}S(\rho)\,=\,\tr\bigl(\rho\hspace*{0.5pt}H\bigr)+T\,\tr\bigl(\rho\ln\rho\bigr)\,.
    \label{eq:non equilibrium free energy appendix}
\end{align}
For~a thermal state, $\tau(\beta)=e^{-\beta H}/\mathcal{Z}$, with the partition function~$\mathcal{Z}\in\mathbb{R}$, and~$\beta=1/T$, where we have set~$k_{\protect\raisebox{-0pt}{\tiny{B}}}=1$, the free energy reduces to
\begin{align}
    F\bigl(\tau(\beta)\bigr)    &=\,-\,T\,\ln\mathcal{Z}\,.
    \label{thermal state free energy}
\end{align}
Moving~a thermal state away from equilibrium always requires work, which is given by the free energy difference
\begin{align}
    \Delta F\bigl(\tau(\beta)\rightarrow\rho\bigr)  &=\,F(\rho)\,-\,F(\tau)
    \label{eq:free energy difference appendix}
\end{align}
to the final state~$\rho$. An elementary way to see that~\mbox{$\Delta F\geq0~\forall~\rho$} for initial thermal states is via the relative entropy~$S(\rho\dvline\tau)$, defined as
\begin{align}
    S(\rho\dvline\tau)   &=\,-S(\rho)\,-\,\tr\bigl(\rho\ln\tau\bigr)\,.
    \label{eq:relative entropy definiton}
\end{align}
For thermal states~$\tau(\beta)$ we may then write
\begin{align}
    T\,S\bigl(\rho\dvline\tau(\beta)\bigr)   &=\,-T\,S(\rho)\,-\,T\,\tr\bigl(\rho\ln\tau(\beta)\bigr)
    \nonumber\\[1mm]
    &\ =\,-T\,S(\rho)\,+\tr(\rho\hspace*{0.5pt}H)\,+\,T\,\ln\mathcal{Z}\,\tr(\rho)
    \nonumber\\[1mm]
    &\ =\,F(\rho)\,-\,F\bigl(\tau(\beta)\bigr)\nonumber\\[1mm]
    &\ =\,\Delta F\bigl(\tau(\beta)\rightarrow\rho\bigr)\,.
    \label{eq:relating rel entropy and free energy}
\end{align}
By virtue of Klein's inequality (see, \eg, Ref.~\cite{Ruskai2002}), the quantum relative entropy is non-negative, $S(\rho\dvline\tau)\geq0$, and vanishes if and only if~$\rho=\tau$. Consequently, we can conclude that~$\Delta F\bigl(\tau(\beta)\rightarrow\rho\bigr)\geq0$.


\section{Energy cost of a general unitary}\label{sec:methods B Energy cost of a general unitary}

We now give a~detailed proof of Eq.~(\ref{eq:free energy difference S vs B}), where we denote the transformed states of the system, the subsystems, and the bath as~$\rho\Sys$, $\rho\Sone$, $\rho\Stwo$, and $\rho\bath$, respectively. Starting from Eq.~(\ref{eq:W in}), the energy differences are rewritten in terms of the changes in free energy and entropy as
\begin{align}
    W   &=\,\Delta E\Sys+\Delta E\bath\nonumber\\[1mm]
    &\ =\,\Delta F\Sys+\Delta F\bath\,+\,T\,\bigl[S(\rho\Sys)+S(\rho\bath)-S(\tau\Sys)-S(\tau\bath)\bigr]
    \nonumber\\[1mm]
    &\ =\,\Delta F\Sys+\Delta F\bath\,+\,T\,\bigl[S(\rho\Sys)+S(\rho\bath)-S(\tau\SB)\bigr]
    \nonumber\\[1mm]
    &\ =\,\Delta F\Sys+\Delta F\bath\,+\,T\,\bigl[S(\rho\Sys)+S(\rho\bath)-S(\rho\SB)\bigr]
    \nonumber\\[1mm]
    &\ =\,\Delta F\Sys+\Delta F\bath\,+\,T\,\mathcal{I}\SB\,,
    \label{eq:free energy difference S vs B detailed proof}
\end{align}
where we have made use of the fact that the global unitary leaves the overall entropy unchanged, $S(\rho\SB)=S(\tau\SB)$. To prove the similar result of Eq.~(\ref{eq:free energy difference S1 vs S2}) for the partition of the system~$S$ into its subsystems we first write
\begin{align}
    \Delta F\Sys    &=\,\Delta E\Sys\,-\,T\,\Delta S\Sys\nonumber\\[1mm]
    &\ =\,\Delta E\Sone+\Delta E\Stwo-T\,\bigl[S(\rho\Sys)-S(\tau\Sys)\bigr]\,.
    \label{eq:free energy difference S1 vs S2 detailed proof part 1}
\end{align}
The energy differences of the subsystems may then be expressed as
\begin{subequations}
\label{eq:free energy difference S1 vs S2 detailed proof part 2}
\begin{align}
    \Delta E\Sone   &\ =\,\Delta F\Sone+T\,\bigl[S(\rho\Sone)-S(\tau\Sone)\bigr]\,,
    \label{eq:free energy difference S1 vs S2 detailed proof part 2 S 1}\\[1mm]
    \Delta E\Stwo   &\ =\,\Delta F\Stwo+T\,\bigl[S(\rho\Stwo)-S(\tau\Stwo)\bigr]\,.
    \label{eq:free energy difference S1 vs S2 detailed proof part 2 S 2}
\end{align}
\end{subequations}
Finally, noting that $S(\tau\Sone)+S(\tau\Stwo)=S(\tau\Sys)$, one arrives at
\begin{align}
    \Delta F\Sys    &=\,\Delta F\Sone+\Delta F\Stwo+T\,\bigl[S(\rho\Sone)+S(\rho\Stwo)-S(\rho\Sys)\bigr]
    \nonumber\\[1mm]
    &\ =\,\Delta F\Sone\,+\,\Delta F\Stwo\,+\,T\,\mathcal{I}\Sonetwo\,,
    \label{eq:free energy difference S1 vs S2 detailed proof 3}
\end{align}
which concludes the proof.


\section{Optimal protocol for generating mutual information}\label{sec:methods C Optimal protocol for generating mutual information}

Let us now turn our attention to the protocol for the optimal generation of correlations. We prove here that the ultimate bound~$W=T\nr\mathcal{I}\Sonetwo$ can be achieved, by first proving Eq.~(\ref{eq:optimal correlation protocol work}). The (minimal) energy cost~$W_{\mathrm{I}}$ for the first step, reducing the system temperature from~$T$ to~$\Tone\leq T$, is given by
\begin{align}
    W_{\mathrm{I}}  &=\,\Delta F\Sys\bigl(\tau\Sys(\beta)\rightarrow\tau\Sys(\bone)\bigr)\,=\,
    E\bigl(\tau\Sys(\bone)\bigr)-E\bigl(\tau\Sys(\beta)\bigr)\nonumber\\[1mm]
    &\ -\,T\,\Bigl[S\bigl(\tau\Sys(\bone)\bigr)-S\bigl(\tau\Sys(\beta)\bigr)\Bigr]\,.
    \label{eq:step I work appendix}
\end{align}
For the second step we use~a unitary operation, which leaves the system entropy invariant, while the subsystems become locally thermal at temperature~$\Ttwo=1/\btwo$. The average energy of the system after the transformation is hence identical to that of~a thermal state~$\tau\Sys(\btwo)$. The minimal energy cost~$W_{\mathrm{I\nl I}}$ is hence given by
\begin{align}
    W_{\mathrm{I\nl I}}  &=\,
    E\bigl(\tau\Sys(\btwo)\bigr)-E\bigl(\tau\Sys(\bone)\bigr)\,.
    \label{eq:step II work appendix}
\end{align}
The correlations of the final state, as measured by the mutual information, are then
\begin{align}
    \mathcal{I}\Sonetwo  &=\,S\bigl(\tau\Sone(\btwo)\bigr)+S\bigl(\tau\Stwo(\btwo)\bigr)-S\bigl(\tau\Sys(\bone)\bigr)
    \nonumber\\[1mm]
    &\ =\,S\bigl(\tau\Sys(\btwo)\bigr)-S\bigl(\tau\Sys(\bone)\bigr)\,.
    \label{eq:step II mutual information appendix}
\end{align}
Using Eq.~(\ref{eq:step II mutual information appendix}), the energy costs for both steps can be combined to arrive at
\begin{align}
    W   &=\,W_{\mathrm{I}}\,+\,W_{\mathrm{I\nl I}}\,=\,E\bigl(\tau\Sys(\btwo)\bigr)-E\bigl(\tau\Sys(\beta)\bigr)\nonumber\\[1mm]
    &\ -\,T\,\Bigl[S\bigl(\tau\Sys(\btwo)\bigr)-S\bigl(\tau\Sys(\beta)\bigr)-\mathcal{I}\Sonetwo\Bigr]\nonumber\\[1mm]
    &=\,\Delta F\Sys\bigl(\tau\Sys(\beta)\rightarrow\tau\Sys(\btwo)\bigr)\,+\,T\,\mathcal{I}\Sonetwo\nonumber\\[1mm]
    &=\,T\,\Bigl[S\bigl(\tau\Sys(\btwo)\dvline\tau\Sys(\beta)\bigr)\,+\,\mathcal{I}\Sonetwo\Bigr]\,.
    \label{eq:optimal correlation protocol work appendix}
\end{align}

Now, if~$W$ is split into the contributions $W_{\mathrm{I}}$ and $W_{\mathrm{I\nl I}}$ such that $\btwo=\beta$, one obtains $T\nr\mathcal{I}\Sonetwo = W$, as desired. Interestingly, this is not always achievable. Setting~$\btwo=\beta$ may require~$W_{\mathrm{I}}$ to become larger than the energy that is necessary to cool down to the ground state. This leads to~a surplus of energy for the correlation step. In such a case, $\Ttwo$ is larger than the initial temperature~$T$. The transition to this regime occurs when,
\begin{align}
    W   &=\tilde{W}=\tilde{W}_{\mathrm{I}}\,+\,\tilde{W}_{\mathrm{I\nl I}}\,=\,T\nr S\bigl(\tau\Sys(\beta)\bigr)\,,
    \label{eq:transition between energy regimes}
\end{align}
where~$\tilde{W}_{\mathrm{I}}=-F\bigl(\tau\Sys(\beta)\bigr)$ corresponds to the energy necessary to cool down to the ground state and~$\tilde{W}_{\mathrm{I\nl I}}=E\bigl(\tau\Sys(\beta)\bigr)$ is the work necessary to correlate the systems such that $\btwo=\beta$. After some rearranging, one obtains
\begin{align}
\mathcal{I}\Sonetwo &\leq
    \begin{cases}
        \beta\nr W     &\ \mbox{if}\ \beta\nr W \leq  S\bigl(\tau\Sys(\beta)\bigr)\,,  \\
        S(\tau\Sys(\btwo)) &\ \mbox{if}\ \beta\nr W >  S\bigl(\tau\Sys(\beta)\bigr)\,,
    \end{cases}
    \label{eq:energy regimes mutual inf appendix}
\end{align}
where $\btwo$ is given by the implicit relation
\begin{align}
    E\bigl(\tau\Sys(\btwo)\bigr)    &=\,W\,+\,F\bigl(\tau\Sys(\beta)\bigr)\,.
    \label{eq:transition beta equal beta two implicit}
\end{align}
There are thus two fundamentally different regimes for the generation of mutual information.


\section{Generation of mutual information between two bosonic modes}\label{sec:methods E Generation of mutual information in two bosonic modes}

Let us examine more closely the scaling of the generated correlations with the input energy. Since the amount of energy that may be used to correlate two fermionic modes is finite, we will focus on the system of two bosonic modes as described in Section~\ref{sec:bosonic systems main}. Recall that the system Hamiltonian is given by $H\Sys=H\Sone+H\Stwo$. Up to~a constant, the subsystem Hamiltonians may be expressed in terms of the Fock states~$\ket{n\SoneorStwo}=(1/\sqrt{n!})(a_{\raisebox{-0.5pt}{\tiny{1(2)}}}^{\dagger})^{n}\ket{0}$ as
\begin{align}
    H\SoneorStwo &=\,\sum\limits_{n=0}^{\infty} n\nr \omega \ket{n\SoneorStwo}\!\bra{n\SoneorStwo}\,,
    \label{eq:harmonic oscillator}
\end{align}
and we use units where $\hbar=1$. Likewise, the initial thermal state~$\tau\Sys(\beta)=\tau\Sone(\beta)\otimes\tau\Stwo(\beta)$ can be expressed in this way, \ie,
\begin{align}
    \tau\SoneorStwo(\beta)  &=\,\sum\limits_{n=0}^{\infty}p_{n}(\beta) \ket{n\SoneorStwo}\!\bra{n\SoneorStwo}\,,
    \label{eq:initialstate}
\end{align}
where $p_{n}=(1-e^{-\beta})e^{-n \beta}$, with~$\beta=1/T$, and temperatures in units of~$\omega$. The energy and entropy of the thermal state evaluates to
\begin{align}
    E\bigl(\tau\Sys(\beta)\bigr)    &=\nr\tr\bigl(H\Sys\tau\Sys(\beta)\bigr)\nr=\nr\omega\nr\Bigl[\coth\bigl(\beta/2\bigr)-1\Bigr]\,,
    \label{eq:Ebetacomparison}\\
    S\bigl(\tau\Sys(\beta)\bigr)    &=\nr-\tr\bigl(\tau\Sys \ln(\tau\Sys)\bigr)\nr=\nr2\nr f\Bigl(\coth\bigl(\beta/2\bigr)\Bigr)\,,
    \label{eq:Sbetacomparison}
\end{align}
where~$f(x)$ is the entropic function
\begin{align}
    f(x)    &=\,\frac{x+1}{2} \ln\Bigl(\frac{x+1}{2}\Bigr)-\frac{x-1}{2} \ln\Bigl(\frac{x-1}{2}\Bigr)\,.
    \label{eq:entropic function}
\end{align}
As we have argued in Eq.~(\ref{eq:energy regimes mutual inf}), the optimal mutual information that may be generated from such~a thermal state using energies~$W$ smaller than~$S\bigl(\tau\Sys(\beta)\bigr)/\beta$ scales linearly with~$W$.

Let us now consider the regime where the supplied energy~$W$ is much larger than~$S\bigl(\tau\Sys(\beta)\bigr)/\beta$. After reaching the ground state in the first step of the protocol, all of the excess energy increases the correlations. The energy of the final state is equal to the work invested into the correlation step, \ie, $E\bigl(\tau\Sys(\btwo)\bigr)=W_{\mathrm{I\nl I}}$. From Eq.~(\ref{eq:Ebetacomparison}), we hence find
\begin{align}
    \coth\bigl(\frac{\btwo}{2}\bigr)\    &=\,\frac{W_{\mathrm{I\nl I}}}{\omega}\,+\,1\,.
    \label{eq:high energy regime work}
\end{align}
From Eq.~(\ref{eq:energy regimes mutual inf appendix}) we infer that the mutual information is given by~$\mathcal{I}\Sonetwo=S(\tau\Sys(\btwo))$. Inserting into Eq.~(\ref{eq:Sbetacomparison}) and expanding~$f\bigl((W_{\mathrm{I\nl I}}/\omega)+1\bigr)$ into~a Taylor-Maclaurin series for~$(\omega/W_{\mathrm{I\nl I}})\ll1$, we find
\begin{align}
    \mathcal{I}\Sonetwo &=\,
    2+2\ln\bigl(\tfrac{1}{2}\frac{W_{\mathrm{I\nl I}}}{\omega}\bigr)+\mathcal{O}\bigl(\frac{\omega}{W_{\mathrm{I\nl I}}}\bigr)\,,
    \label{eq:high energies mut inf scaling proof}
\end{align}
where $\mathcal{O}(x)$ is~a quantity such that $\mathcal{O}(x)/x$ remains finite in the limit~$x\rightarrow0$. We conclude that for large energy supply, the optimally generated correlations increase only logarithmically with increasing energy, in stark contrast to the linear increase at small energies, see Fig.~\ref{fig:bosons optimal EOF with bath}~(a).


\section{Optimal protocol for fermionic entanglement of formation}\label{sec:Optimal protocol for fermionic entanglement of formation}

We now present~a modification of our previous protocol for the generation of entanglement between two fermionic modes. To optimally convert the supplied energy into fermionic entanglement of formation, the temperatures of the two modes are allowed to change independently of each other in the first step of the protocol. In particular, this entails heating as well as cooling of the individual modes, and the average particle numbers~$N^{\mathrm{I}}\Sone$ and~$N^{\mathrm{I}}\Stwo$ may be different from each other. As before, the energy cost~$W_{\mathrm{I}}$ for this step is given by the free energy difference of the initial thermal and the transformed state.

For step~$\mathrm{I\nl I}$ of the protocol, the two modes are correlated using unitary operations on the system only. As mentioned before, the superselection rules forbid coherent superpositions between even and odd numbers of fermions. In particular, the maximally entangled two-mode pure states for the even parity subspace, $\fket{\!\phi^{\pm}\!}=\frac{1}{\sqrt{2}}\bigl(\fket{\!0\!}\pm\fket{\!1\Sone\!}\!\fket{\!1\Stwo\!\!}\bigr)$, and those for the odd parity subspace,
\mbox{$\fket{\!\psi^{\pm}\!}=\frac{1}{\sqrt{2}}\bigl(\fket{\!1\Sone\!}\pm\fket{\!1\Stwo\!\!}\bigr)$}, may not be interconverted by parity conserving operations.
These states may hence be regarded as forming~a maximally entangled set~\cite{DeVicenteSpeeKraus2013}.
Consequently, the optimally correlating unitary $U_{\mathrm{corr}}$ for two modes decomposes into~a direct sum of two~$SU(2)$ rotations. For each, only one real parameter, denoted by~$\theta_{\mathrm{even}}$ and~$\theta_{\mathrm{odd}}$, respectively, is relevant for the amount of generated entanglement. We quantify the entanglement by the superselected entanglement of formation, \ie, the minimum number, per copy, of the aforementioned maximally entangled states respecting superselection rules, that are required to assemble~a given two-mode state.

However, note that the imposed superselection rules also prevent local changes of basis for each fermionic mode. The states $\fket{\!\phi^{\pm}\!}$ and~$\fket{\!\psi^{\pm}\!}$
could therefore be considered to be entangled only in~a mathematical sense, that is, the entanglement may not be directly used, for instance, to violate~a Bell inequality. Nonetheless, if the entanglement is extracted by swapping it to~a bosonic system, it becomes useful in the conventional sense. Since~a swap using local unitaries cannot create entanglement, its origin must lie in the original fermionic entanglement. Keeping this argument in mind,~a pure state decomposition of the transformed state that requires the fewest copies of the maximally entangled pure states $\fket{\!\phi^{\pm}\!}$ and~$\fket{\!\psi^{\pm}\!}$ may easily be found, yielding the entanglement of formation
\begin{align}
    E_{oF}
    &=\,\ln(2)\bigl[|1-N^{\mathrm{I}}\Sone-N^{\mathrm{I}}\Stwo|\sin(2\theta_{\mathrm{even}})\nonumber\\[1mm]
    &\ +\,|N^{\mathrm{I}}\Sone-N^{\mathrm{I}}\Stwo|\sin(2\theta_{\mathrm{odd}})\bigr]\,,
    \label{eq:EoF fermions}
\end{align}
where~$0\leq\theta_{\mathrm{even}},\theta_{\mathrm{odd}}\leq\pi/4$. Since the odd-subspace rotation shifts excitations of equal frequency,~$\theta_{\mathrm{odd}}$ does not contribute to the energy cost of the second step, which is given by
\begin{align}
    \frac{W_{\mathrm{I\hspace*{-0.5pt}I}}}{\omega}    &=\,
    2\bigl(1-N^{\mathrm{I}}\Sone-N^{\mathrm{I}}\Stwo\bigr)\sin^{2}(\theta_{\mathrm{even}})\,.
    \label{eq:fermions energy cost opt entanglement step 2}
\end{align}
We may hence set~$\theta_{\mathrm{odd}}=\pi/4$ at no additional expense in energy. We note that this suggests~a tradeoff between creating entanglement in the even and odd subspace by heating one mode, while the other is cooled. The entanglement of formation becomes maximal when enough energy is supplied to cool one mode, we assume here~$S_{1}$, to the ground state, while $\theta_{\mathrm{even}}=\tfrac{\pi}{4}$. The minimum energy~$W_{\mathrm{opt}}$ for which this is the case is obtained when the reduced state of the second mode~$S_{2}$ is maximally mixed. If less energy than~$W_{\mathrm{opt}}$ is supplied, it is split between cooling and heating the modes~$S_{1}$ and~$S_{2}$, respectively, in step~$\mathrm{I}$, before correlating them in step~$\mathrm{I\hspace*{-0.5pt}I}$. The resulting state is~a mixed state that is entangled both in the even and odd subspace. When $W=W_{\mathrm{opt}}$, the weights of the even and odd subspace entangled states are equal.

\begin{figure}[ht!]
(a)\includegraphics[width=0.47\textwidth]{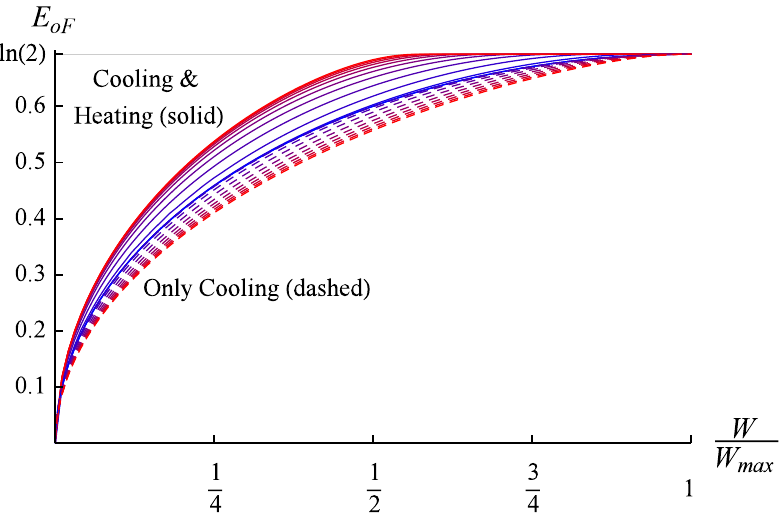}
(b)\includegraphics[width=0.47\textwidth]{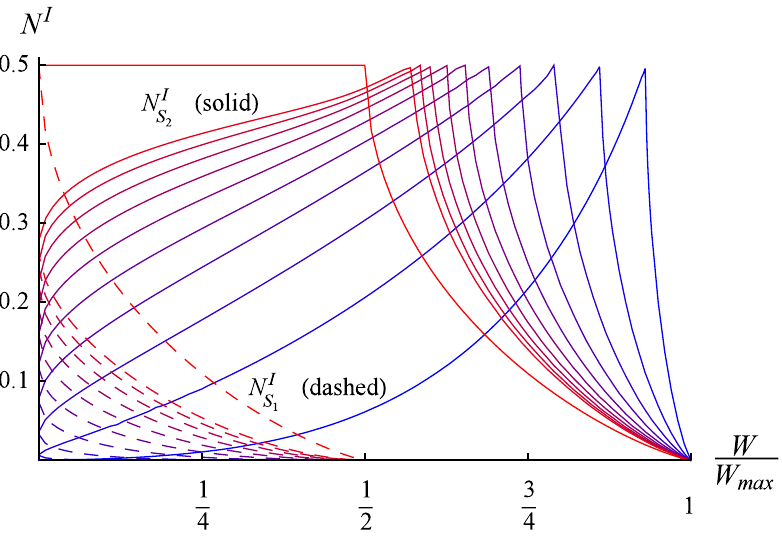}
\caption{\label{fig:fermions optimal EOF with bath}
\textbf{Optimal fermionic entanglement}: The solid curves in Fig.~\ref{fig:fermions optimal EOF with bath}~(a) show the amount of entanglement (of formation) that can maximally be generated between two fermionic modes,~$S_{1}$ and~$S_{2}$, that are initially in~a thermal state of temperature~$T$, for~a given energy cost~$W$.  The horizontal axis shows the relative energy cost, \ie, the fraction of~$W$ and the energy cost~$W_{\mathrm{max}}$. Fig.~\ref{fig:fermions optimal EOF with bath}~(b) shows the average particle numbers~$N\Sone^{\mathrm{I}}$ (dashed lines) and~$N\Stwo^{\mathrm{I}}$ (solid lines) after the first step of the protocol, where we have assumed $N\Stwo^{\mathrm{I}}\geq N\Sone^{\mathrm{I}}$ without loss of generality. The curves in both Fig.~\ref{fig:fermions optimal EOF with bath}~(a) and~(b) are plotted for initial temperatures varying from $T=0$ to $T=1$ in steps of $0.1$ and for the limit~$T\rightarrow\infty$ (bottom to top) in units~$\hbar\omega/k_{\protect\raisebox{-0pt}{\tiny{B}}}$. The dashed curves in Fig.~\ref{fig:fermions optimal EOF with bath}~(a) show the corresponding curves of Fig.~\ref{fig:fermions optimal even entanglement with bath}~(a) for temperatures varying from $T=0$ to $T=1$ in steps of $0.1$ (top to bottom) as a comparison.}
\end{figure}

As more energy is provided, it may be used to shift the entanglement to one of the subspaces, obtaining~a final state with higher purity. When~$W=W_{\mathrm{max}}$, where $W_{\mathrm{max}} =W_{\mathrm{opt}}+T\ln(2)=2T\ln(e^{\beta}+1)-\omega$, the final overall state is pure, but the entropy of both subsystems is maximal. The exact values of~$N^{\mathrm{I}}\Sone$, $N^{\mathrm{I}}\Stwo$, and $\theta_{\mathrm{even}}$ may be determined by numerical optimization for fixed values of~$W$ and~$T$. In Fig.~\ref{fig:fermions optimal EOF with bath}, the protocol is illustrated for various temperatures, where the excess energy between~$W_{\mathrm{opt}}$ and~$W_{\mathrm{max}}$ is used to shift the entanglement towards the even subspace.

Note that the single-mode marginals of the superselected fermionic modes after step~$\mathrm{I}$ of the protocol are fully determined by the corresponding average particle numbers. In principle, one may therefore consider the first step to involve the preparation of more general, uncorrelated states, for which~$1/2<N^{\mathrm{I}}\Soneortwo\leq1$. However, we find that optimality is achieved for particles numbers that are compatible with thermal marginals.

\section{Optimality of Gaussian operations}\label{sec:methods E About the optimality of gaussian operations}

Finally, we investigate the optimality of Gaussian operations for the generation of entanglement. As for the mutual information, we identify two energy regimes with qualitatively different behavior. In~a certain low-energy regime, we are able to show that Gaussian operations are not optimal. To achieve this, we construct~a protocol using specific non-Gaussian unitaries, which outperforms our previously established protocol for Gaussian operations. Nevertheless, in the high-energy regime, Gaussian operations perform better. Indeed, we show that the entanglement generated by the Gaussian protocol scales optimally with the available energy in this case.



\subsection*{Low-energy regime}

\begin{figure}[ht!]
\label{fig:Comparing}
\includegraphics[width=0.51\textwidth]{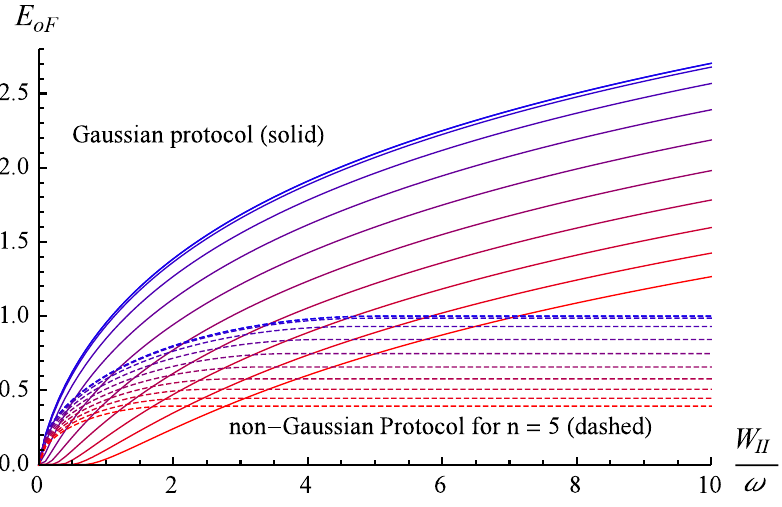}
\caption{
\textbf{Comparison of Gaussian and non-Gaussian operations}:
The plot shows the amount of entanglement (of formation) that can maximally be generated between two bosonic modes in step~$\mathrm{I\nl I}$ of the protocol. Both modes are assumed to have been cooled to temperature~$\Tone$ in the first step. Using the energy~$W_{\mathrm{I\nl I}}$, the solid curves show the optimal entanglement generated by Gaussian operations, while the dashed curves show the amount of entanglement generated by the non-Gaussian protocol. In both cases, the curves are plotted for temperatures varying from $\Tone=0$ to $\Tone=1$ in steps of $0.1$  (top to bottom) in units~$\hbar\omega/k_{\protect\raisebox{-0pt}{\tiny{B}}}$.}
\end{figure}

Instead of the previously established protocol based on Gaussian operations, we now introduce~a scheme to generate entanglement using non-Gaussian operations in the correlation step. That is, after cooling the system to the temperature~$\Tone=1/\bone$ using the energy~$W_{\mathrm{I}}$, we perform~a unitary transformation that rotates in the subspace of the two-mode Fock space that is spanned by $\ket{0\Sone}\ket{0\Sone}$ and $\ket{n\Sone}\ket{n\Stwo}$, where we recall the notation of Appendix~\ref{sec:methods E Generation of mutual information in two bosonic modes}. One may think of this operation as generating Bell states in the four-dimensional subspace. We conveniently parametrize this rotation by~a single, real parameter~$\alpha$, where~$0\leq\alpha\leq\pi/4$, such that
\begin{small}
\begin{align}
    \ket{0\Sone}\!\ket{0\Stwo}    &\mapsto\, \cos(\alpha)\ket{0\Sone}\!\ket{0\Stwo}+\sin(\alpha)\ket{n\Sone}\!\ket{n\Stwo}\,,
    \label{eq:Bell state rotation}\\
    \ket{n\Sone}\!\ket{n\Stwo}    &\mapsto\, \cos(\alpha)\ket{n\Sone}\!\ket{n\Stwo}-\sin(\alpha)\ket{0\Sone}\!\ket{0\Stwo}\,.
\end{align}
\end{small}
The energy cost~$W_{\mathrm{I\nl I}}$ of this rotation is given by
\begin{align}
    W_{\mathrm{I\nl I}} &=\,2\nr n\nr\omega\nr\bigl(p_{0}^{2}-p_{n}^{2}\bigr)\nr\sin^{2}(\alpha)\,,
    \label{eq:non-Gaussian energy cost}
\end{align}
where we now have $p_{n}=(1-e^{-\bone})e^{-n \bone}$, with~$\bone=1/\Tone$, and temperatures in units of~$\omega$. Here, the entanglement of formation of the transformed state can be quantified by way of the concurrence\cite{Wootters1998} of the (unnormalized) state of the subspace spanned by $\ket{0\Sone}\ket{0\Sone}$, $\ket{0\Sone}\ket{n\Sone}$, $\ket{n\Sone}\ket{0\Stwo}$, and $\ket{n\Sone}\ket{n\Stwo}$, see Refs.~\cite{HashemiRafsanjaniHuberBroadbentEberly2012,HuberDeVicente2013}. For the concurrence~$\mathcal{C}$, we obtain the expression
\begin{align}
    \mathcal{C} &=\,\bigl(p_{0}^{2}-p_{n}^{2}\bigr)\sin(2\alpha)\,-\,2\nr p_{0}\nr p_{n}
    \label{eq:concurrence non-Gaussian protocol}\\
        &=\,
        \sqrt{\tfrac{1}{n}\tfrac{W_{\mathrm{I\nl I}}}{\omega}}
        \sqrt{2\bigl(p_{0}^{2}-p_{n}^{2}\bigr)-\tfrac{1}{n}\tfrac{W_{\mathrm{I\nl I}}}{\omega}}
        -2\nr p_{0}\nr p_{n}\,.\nonumber
\end{align}
Whenever~$\mathcal{C}>0$, entanglement is present, which translates to the condition
\begin{align}
    \frac{W_{\mathrm{I\nl I}}}{\omega} \Bigl(p_{0}^{2} - p_{n}^{2}-\frac{1}{2n}\frac{W_{\mathrm{I\nl I}}}{\omega}\Bigr)
    &>\, 2\nr n\nr p_{0}^{2}\nr p_{n}^{2}\,.
    \label{eq:non-Gaussian ent condition}
    \vspace*{-2mm}
\end{align}
It can easily be seen that this condition can always be satisfied by choosing~$n$ to be large enough. Therefore,  \emph{some} entanglement can be generated at an arbitrarily low energy cost given two infinite-dimensional systems. Recall that Gaussian operations require at least the energy~$\frac{\omega}{2}(\nu^{\mathrm{I}}-1)^{2}$ to leave the separable set. Consequently, Gaussian operations cannot be optimal for entanglement generation in all regimes, although they are optimal for the generation of total correlations. Specifically, the unitary of Eq.~(\ref{eq:thermal marginals step II}) can be implemented with Gaussian operations. On the other hand, the amount of entanglement generated by the non-Gaussian protocol we have presented here is bounded. For fixed~$n$, the maximal amount of energy useful for this protocol is~$n\bigl(p_{0}^{2}-p_{n}^{2}\bigr)$, and the corresponding maximal concurrence is given by
\vspace*{-1.5mm}
\begin{align}
   \mathcal{C}_{\mathrm{max}}   &=\,\bigl(p_{0}-p_{n}\bigr)^{2}\,.
\end{align}
In contrast, the entanglement that may be generated by Gaussian operations is unbounded. Our considerations are illustrated in Fig.~\ref{fig:Comparing}.

\subsection*{High-energy regime}

To study the regime of large energies, we first show that Gaussian operations are optimal to generate entanglement from the ground state. If the state is pure, the entanglement of formation is simply given by the entropy of the local state. For a given amount of work, the unitary maximizing $E_{oF}$ will then be precisely the expression of Eq.~(\ref{eq:thermal marginals step II}), as the thermal state maximizes the entropy for~a given energy. Given two bosonic modes, this operation can be implemented by~a two-mode squeezing operation. In the protocols that we have considered, the first step consists of cooling. Whenever the ground state is reached, the Gaussian correlating operation is optimal. This occurs when $W_{\mathrm{I}} > -F\bigl(\tau\Sys(\beta)\bigr)$, and we conclude that the protocol is certainly optimal when~$W\gg-F\bigl(\tau\Sys(\beta)\bigr)$.

\end{document}